\documentclass[letterpaper,final]{article}
\usepackage{osajnl2} 

\begin{document}

\title{Effect of shell thickness on exciton and biexciton binding energy of a ZnSe/ZnS core/shell quantum dot}

\author{Pratima Sen,$^{*}$ Saikat Chattopadhyay, J. T. Andrews$^1$ and P. K. Sen$^{1}$}
\address{Laser Bhawan, School of Physics, \\ Devi Ahilya University, Indore (M.P.), India–-452017}
\address{$^1$Department of Applied Physics, Shri. G. S. Institute of Technology \& Science, \\ 23,Park Road, Indore (M.P.), India–-452003}
\address{$^*$Corresponding author: pratimasen@gmail.com}

\begin{abstract}
The exciton and biexciton binding energy have been studied for a ZnSe/ZnS core/shell quantum dot using WKB (Wentzel-Kramers-Brillouin) approximation. The exciton binding energy increases for small shell thickness and for large thickness, the binding energy again starts decreasing. A similar result is obtained for biexcitons where for thicker shells, the biexciton attains antibonding.
\end{abstract}


\maketitle 

\section{Introduction}
Semiconductor Quantum dots (QDs) have drawn significant attention for last couple of decades as they play the role of key elements for modern optoelectronic and spintronic nanodevices. In the semiconductor quantum dots the electrons and holes are confined within the width of a semiconductor layer of low energy gap which is surrounded by another semiconductor which has higher band gap and is lattice matched with the former. The confinement of electrons and holes give rise to novel optical and electronic properties. Such nanostructures are also called as core/shell quantum dots (CSQD). It has been experimentally observed that core/shell quantum dots (CSQD) exhibit improved photoluminescence (PL) efficiency over the bare quantum dot and the thickness of the shell provides further control on optical and electronic properties of QDs. Some reported example of CSQD are CdSe/ZnS, ZnS/CdS, CdSe/HgS, CdSe/CdS, ZnSe/ZnS, etc.

Lad and Mahanuni observed \cite{lad} nearly 42\% increase in photoluminescence (PL) intensity due to the presence of ZnS shell on ZnSe quantum dot. The electronic and optical properties of ZnSe/ZnS and ZnS/ZnSe CSQD were theoretically studied by Goswami et al. \cite{biplab}, where they found that a charge transfer from core to shell takes place and the core shell structure will exhibit a red shift in the absorption spectra. Gorwth and PL study of ZnSe QD was also reported by Chang et al. \cite{chang}. They applied a flow controlled method for growing ZnTe QD embedded in ZnS and found that for shorter growth time, the QD is smaller and the blue shift in the PL spectra is larger. The smaller QD size denote smaller core radius which corresponds to larger confinement potential which gives rise to blue shift in absorption spectra.

The stimulus for the present work stems from the report of the red shift in the PL spectra as well as the increase in the PL intensity in CSQD due to the presence of shell. In our earlier work \cite{thomas,sen1,bafna,puri}, we reported nonlinear and coherent transient optical properties of the single quantum dot. The energy level structure in bare QDs was also reported by our group \cite{jiten} as well as other groups \cite{efros1,efros2,davies,koch} by taking into account the parabolic, spherical, square well and other types of confinement potential functions. In these models, the tunneling of the single particles to the shell was not given due consideration. In our opinion the inclusion of tunneling property should be incorporated to take into account the effect of shell on the energy level structures of core shell QDs. The present paper aims to study the effect of shell thickness on exciton and biexciton binding energies. We have also examined the effect of shell thickness on optical absorption on ZnSe/ZnS QD. In the present paper we have calculated the exciton, biexciton binding energies in a CSQD and have analysed linear absorption in a ZnSe/ZnS core shell quantum dot. The theoretical results exhibit an increase in the binding energies of exciton and biexcitons due to the presence of the shell. Consequently, the transition energies will show a red shift as observed by Mahamuni et al. \cite{lad} in the PL spectra of CSQD.

\section{THEORETICAL FORMULATION}

There are two types of CSQD. In type-I CSQD, the band offset are such that both electron and hole are confined within the core while, in type-II CSQD, only one of the carriers is confined in the core region while, the other carrier may extend to the shell region \cite{kim}. In ZnSe/ZnS CSQD, ZnS provides a strong confinement to the electron and hole of the core region. It has a higher band gap (3.68 eV \cite{neel}) than ZnSe (2.8 eV \cite{chadi}). The valance-band offset ($V_{v}$) between ZnSe/ZnS is $\approx  0.58$ eV and the conduction-band offset ($V_{c}$) is $\approx 0.03$ eV. Hence, there is hardly any conduction-band offset for electrons in ZnSe/ZnS and the electrons are loosely bound. Also, the hole is much heavier in the core i.e. ZnSe ($m^*_h= 0.60m_0$) compared to the hole inside the shell i.e. ZnS ($m^*_h= 0.49m_0$). Consequently, in a ZnSe/ZnS CSQD system the electron wave function can spread over the shell as well as core domain but the hole is preferentially localized inside the core area \cite{mahamuni}. The electron confinement in this type of QD is possible for very small QD structures where both Coulomb as well as confinement potentials are responsible for the binding of the electron--hole pairs.

\begin{figure}[htb]
	\centerline{
\includegraphics[width=8.3cm]{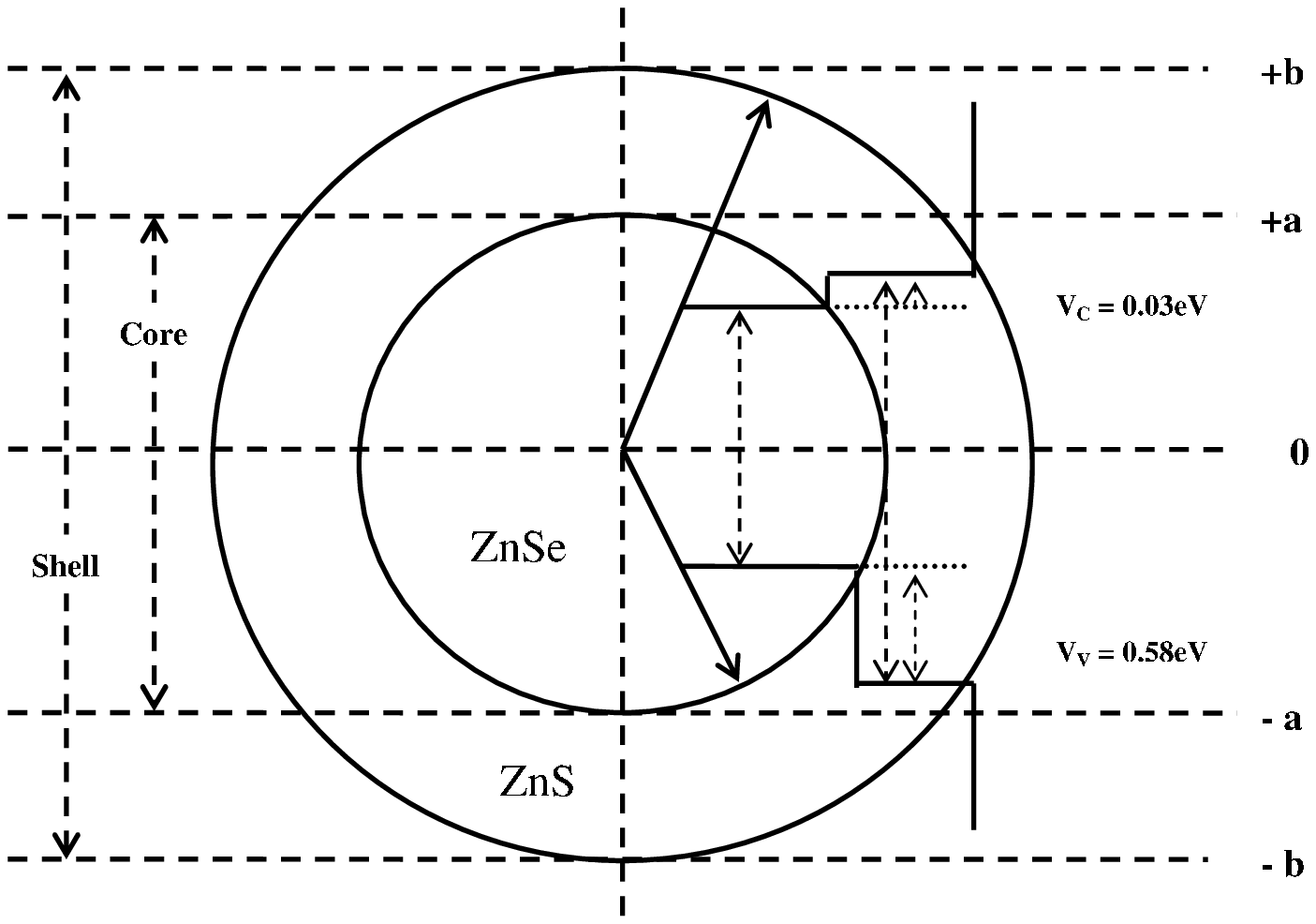}}
	\caption{Schematic of a core-shell quantum dot and dimensions as assumed in the present calculations.}
	\label{sche}
\end{figure}

The geometry of the ZnSe/ZnS QD under consideration is shown in Fig. \ref{sche}. Here $a$ is the radius of the core, while $b$ is the radius of the CSQD as a whole.

\begin{figure}[htb]
\centerline{
\includegraphics[width=8.3cm]{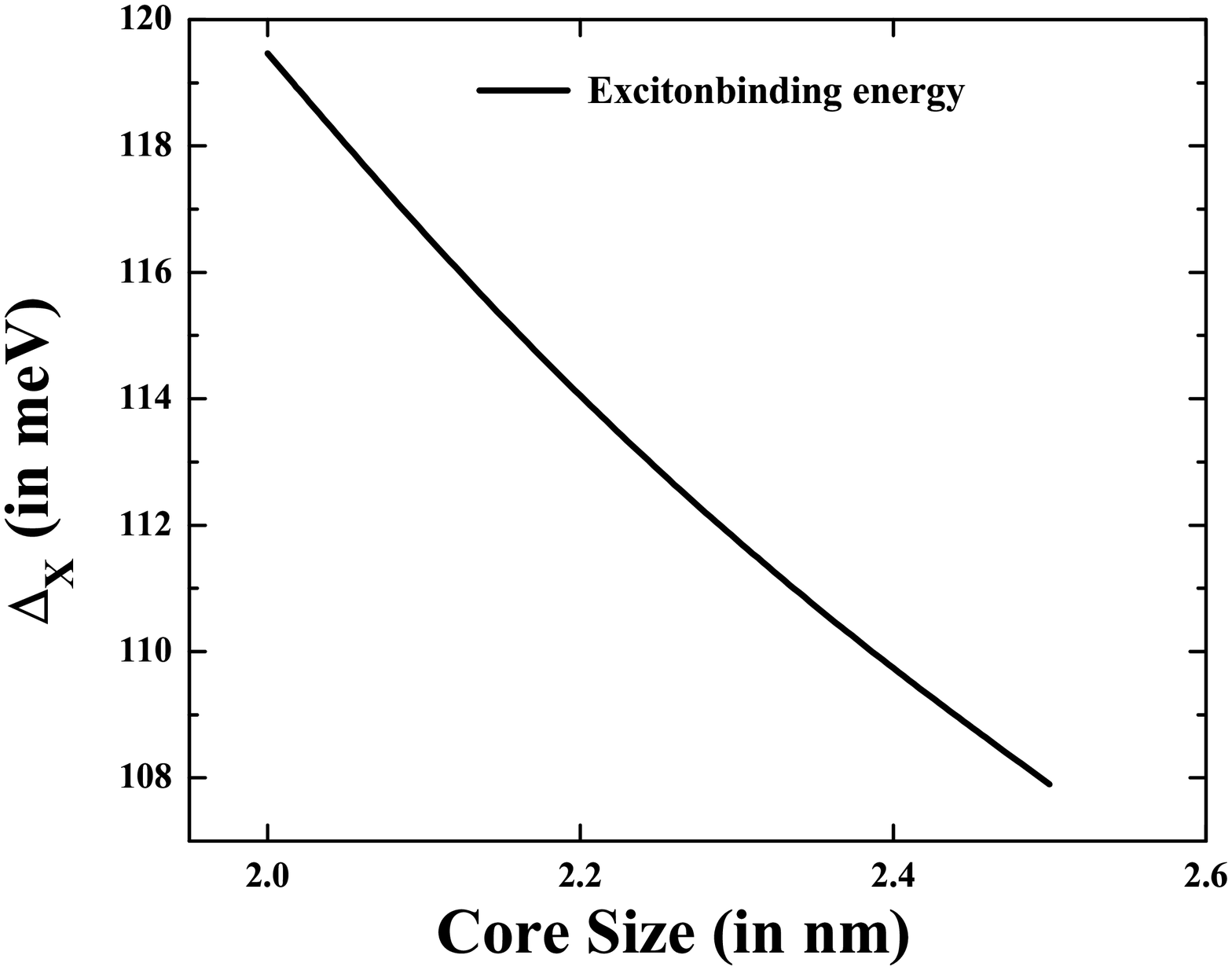}}
\caption{Exciton binding energy with increasing core radius in a ZnSe/ZnS QD.}%
\label{exc}%
\end{figure}

We consider two dimensional confinement potential for electrons $\left(V_{e}\right)$ and holes $\left(V_{h}\right)$ to be parabolic type such that for $\left|x\right|\leq a$, $\left|y\right|\leq a$
\begin{equation}
	V_{e}(x,y)=\frac{V_{c}}{a^{2}}\left(x^2,y^2-a^2\right),
	\label{vce}
\end{equation}
and
\begin{equation}
	V_{h}(x,y)=\frac{V_v}{a^2}\left(x^2,y^2-a^2\right).
	\label{vch}
\end{equation}
Looking at the symmetry of the potential function, in the following discussion only $x$ component is taken into account. We consider that, outside the core region, the particles do not experience strong confinement due to the shell but the particle wave functions still remain weakly confined in the shell region due to the presence of a buffer layer. Consequently, the particle experiences an overall double confinement like structure. The single particle wave functions under such situation can be described by the WKB wave functions \cite{ajoy} given by

\begin{equation}
\phi_{j}\left(a,b\right)=\left\{\begin{array}{l c c} \frac{A_{S}}{\sqrt{\kappa_{j}}}\exp\left[\int^{-a}_{-b}\kappa_{j}dx\right] &\hbox{for } & -b<x<-a, \\
&& \\
\frac{A_{C}}{\sqrt{k_j}}\sin\left[\int^{a}	_{-a}k_{j} dx + \frac{\pi}{4}\right] & \hbox{for } & -a<x<a,  \\
&&\\
\frac{A_{S}}{\sqrt{\kappa_j}}\exp\left[-\int^{b}_{a}\kappa_{j}dx\right] &\hbox{for } & a<x<b. \\
\end{array}\right.
\label{core}
\end{equation}
Here, $j$ stands for electron $(e)$ or hole $(h)$ single particles. $A_{C}$ and $A_{S}$ are the normalization constants for core and shell region respectively, 
\begin{equation}
k_{j}=\left[1-\frac{1}{\hbar}\left(m_{j}V{j}\right)^\frac{1}{2}x\right]^\frac{1}{2} \hbox {and}\;\; \kappa_{j}=\iota k_{j}.
\end{equation}
For a bare quantum dot we restrict the value of the running parameter $x$, to $-a<x<+a$ while the contribution of shell is obtained by giving appropriate values to the running parameter $x$ as $-b<x<-a$ and $a<x<b$ in the forthcoming calculations. Apart from this, the change in the physical parameters regarding energy and effective mass of electron have been approximately incorporated in the core and shell region. The WKB wave functions defined by Eq.\ref{core} are the single particle envelop function. Under effective mass approximation the total wave function $\psi_e$ and $\psi_h$ can be written as the product of the envelop function and the Bloch functions ($u_e,u_h$).

\begin{equation}
\psi_e(a,b,r_e)= \phi_e(a,b)\cdot u_e(r_e),
\label{xy}
\end{equation}
\begin{equation}
\psi_h(a,b,r_h)= \phi_h(a,b)\cdot u_h(r_h).
\label{xy1}
\end{equation}
The optical properties of the semiconductor quantum dots are determined by the electron-hole (e-h) pair states called as, exciton. In QDs the e--h pair formation is influenced by the confinement potential in additional to the Couombic term. Within Hartree approximation \cite{om} the exciton ground state wave function $\left(\psi_X\right)$ can be written as 
\begin{equation}
\centering
\psi_{X}(a,b,r_e,r_h)=\phi_{e}(a,b)u_e(r_e)\cdot\phi_{h}(a,b)u_h(r_h).
\label{xy2}
\end{equation}
Using these wave functions, the exciton binding energy $\Delta_{X}$ can be calculated as \cite{koch}
\begin{eqnarray}
\Delta_{X}&=&\left\langle\psi_{X}\left|\left(V_{e}+\frac{e^2}{\epsilon_{0}x}\right)\right|\psi_{X}\right\rangle  +\left\langle\psi_{X}\left|\left(V_{h}+\frac{e^2}{\epsilon_{0}x}\right)\right|\psi_{X}\right\rangle.
\label{exb}
\end{eqnarray}
In Fig.\ref{exc} we have plotted the variation of exciton binding energy as a function of dot radius in a bare ZnSe QD using Eq. \ref{core}. The physical parameters chosen for calculation of exciton and biexciton binding energies include; the effective mass of electron in ZnSe and ZnS as \cite{dtf,lev} $m_{e}^{*}=0.17m_{0}$ and $m_{e}^{*}=0.39m_{0}$, respectively. The figure exhibits a decrease in exciton binding energy with increasing dot size which is similar to the experimental observations of Chang et. al. \cite{chang} where they calculated the binding energy from the activation energy.

\begin{figure}[htb]
	\centerline{
\includegraphics[width=8.3cm]{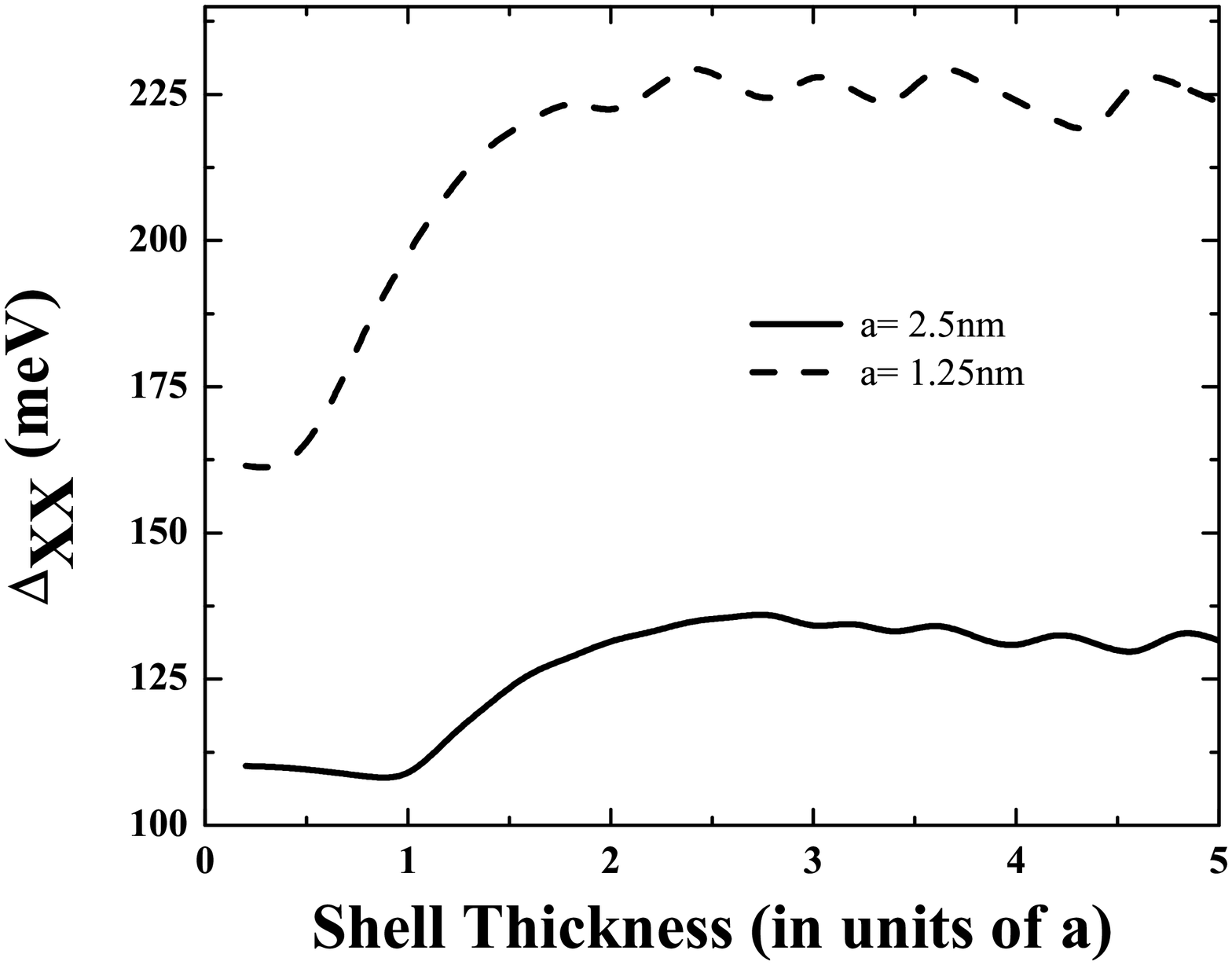}}
\caption{Exciton binding energy with increasing shell thickness in ZnSe/ZnS CSQD.}%
\label{ex}%
\end{figure}

In order to verify the effect of the shell formation on the binding energy of excitons, we have applied our analysis to two quantum dots of the core radius $2.5nm$ and $1.25nm$. For each of the QD the shell width $\left|(b-a)\right|$, is varied from one to two monolayers $(2.5$ x $10^{-10}m)$ to $(5.0 $ x $10^{-10}m)$ to nearly $6$ times the core radius. Fig.\ref{ex} shows the variation of exciton binding energy as a function of shell width and it suggests that the exciton binding energy increases up to few mono layer thickness of the shell width, beyond it starts decreasing. The decreasing trend of exciton binding energy is sharper in large core radii of the QD while it slowly decreases for small core radii QD.

While examining the effect of ZnS shell thickness on confined energy levels of ZnSe QD, Lad and Mahamuni \cite{lad} observed a red shift in the lowest excitonic transition with increase in ZnS shell thickness. The larger red shift shows increase in exciton binding energy which causes reduction in ground to exciton level transition energy. The similarity between our and Lad and Mahamuni's result confirm that the cause of red shift in absorption spectra indicates spreading of single particle wave function to the shell region.

We have further analyzed the effect of shell thickness on biexciton binding energy in a CSQD. The expression for the biexciton binding energy $\Delta_{XX}$ is calculated as \cite{mann}

\begin{eqnarray}
\Delta_{XX}&=&
\left\langle \psi_{e}\left|V_{e}+\frac{e^{2}}{\epsilon_0 x}\right|\psi_{e}\right\rangle  -\left\langle \psi_{h}\left|V_{h}+\frac{e^2}{\epsilon_0 x}\right|\psi_{h}\right\rangle \nonumber\\
&&+\left\langle\psi_{e}\left|V_{h}+\frac{e^2}{\epsilon_0 x}\right|\psi_{e}\right\rangle  -\left\langle\psi_{h}\left|V_{e}+\frac{e^2}{\epsilon_0 x}\right|\psi_{h}\right\rangle.
\label{bxeb}
\end{eqnarray}

\begin{figure}[htb]
\centerline{
\includegraphics[width=8.3cm]{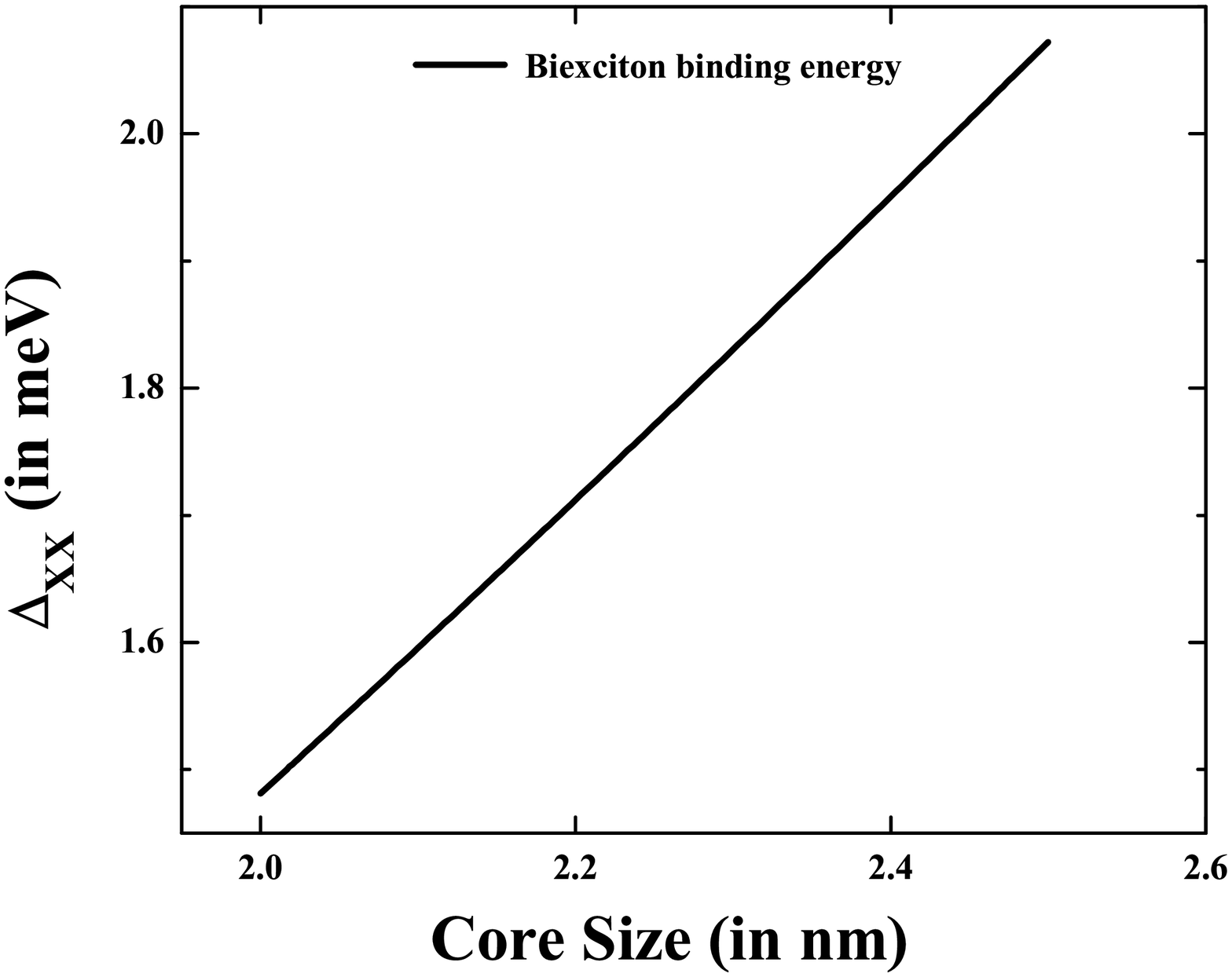}}
\caption{Biexciton binding energy with increasing core radius in a ZnSe/ZnS QD.}%
\label{bxc}%
\end{figure}

\begin{figure}[htb]
	\centerline{
\includegraphics[width=8.3cm]{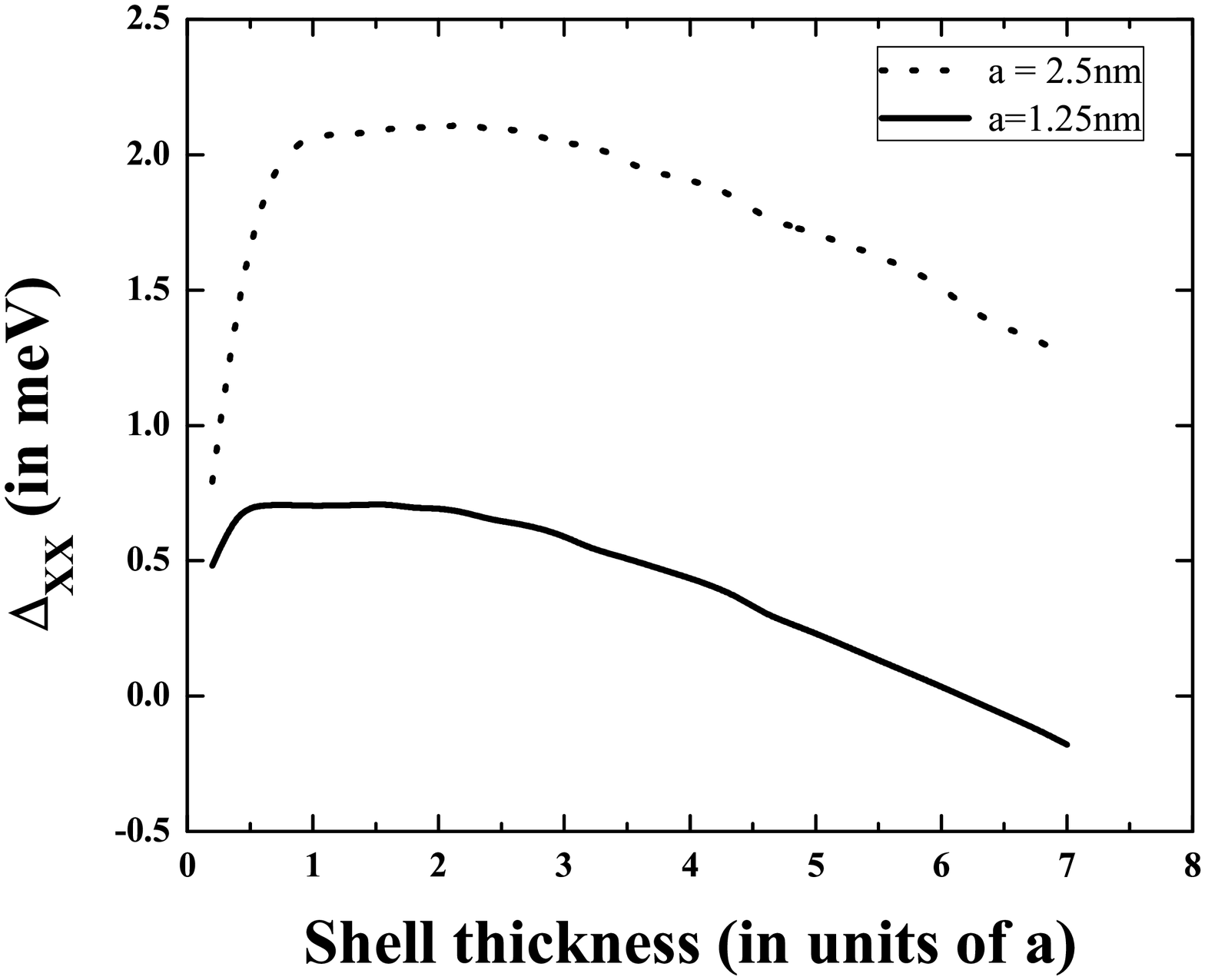}}
\caption{Biexciton binding energy with increasing shell thickness in ZnSe/ZnS CSQD.}
\label{bix}
\end{figure}

The positive (negative) values of $\Delta_{XX}$ indicate bonding (antibonding) biexcitons \cite{mann}.
The calculations were made for a bare as well as a CSQD. 
The variation in biexciton binding energy as a function of dot radius has been plotted in Fig.\ref{bxc}. The figure indicates that, in a bare QD the biexciton binding energy decreases with decreasing dot size, while studying fine structure splitting and biexciton binding energy in single self assembled InAs/AlAs QD, Sarkar et. al. \cite{sarkar} also experimentally observed similar variation. The reduction in the binding energy reflects a possible crossover to the antibonding regime. The influence of shell thickness for core radii $2.5nm$ on biexciton binding energy has been plotted in Fig.\ref{bix} respectively. We find that the biexciton binding energy decreases sharply with increasing shell thickness. In QD of smaller core radii $(1.25nm)$ the biexciton attains an antibonding state beyond a shell width of $6nm$. The antibonding biexciton state occurs due to the fact that the hole wave functions spreads in the shell region in case of small QD. Also the correlation energy decreases due to larger electronic confinement potential of the small core radii QDs. This leads to an enhancement in the last term represented in eq. (\ref{bxeb}) and gives rise to antibonding state of biexciton\cite{rodt}.

\section{CONCLUSIONS}
In conclusion, we find that in a conventional core shell semiconductor QD, the electron and hole wave functions are modified inside the core area due to the presence of the shell layer. WKB approximation method is appropriate to select the modified wave function for the electrons and holes as it explains the experimental observations of increase in exciton binding energy due to the presence of the shell. The theory also successfully explains the bonding and antibonding states of biexcitons in core/shell quantum dot.

\section*{Acknowledgments}
The authors (PS, SC, PKS) thankfully acknowledge the financial support received from Department of Science \& Technology, New Delhi, India.

\clearpage

\listoffigures

\clearpage

\clearpage

\clearpage

\clearpage

\clearpage

\end{document}